# An Experimentally-Validated 3D Electrochemical Model Revealing Electrode Manufacturing Parameters' Effects on Battery Performance

Chaoyue Liu[1,2], Teo Lombardo[1,2], Jiahui Xu[1,2], Alain C. Ngandjong[1,2], Alejandro A. Franco[1,2,3,4*]

[1]Laboratoire de Réactivité et Chimie des Solides (LRCS), CNRS UMR 7314, Université de Picardie Jules Verne, Hub de l'Energie, 15 rue Baudelocque, 80039 Amiens Cedex, France

[2] Réseau sur le Stockage Electrochimique de l'Energie (RS2E), Fédération de Recherche CNRS 3459, Hub de l'Energie, 15 rue Baudelocque, 80039 Amiens Cedex, France

[3]ALISTORE-European Research Institute, Fédération de Recherche CNRS 3104, Hub de l'Energie, 15 rue Baudelocque, 80039 Amiens Cedex, France

[4]Institut Universitaire de France, 103 Boulevard Saint Michel, 75005 Paris, France

*Corresponding author: <u>alejandro.franco@u-picardie.fr</u>

**Abstract**

Electrode manufacturing is at the core of the lithium ion battery (LIB) fabrication process. The electrode microstructure and the electrochemical performance are determined by the adopted manufacturing parameters. However, in view of the strong interdependencies between these parameters, evaluating their influence on the performance is not a trivial task. In our previous publications, we have reported a series of computational models simulating the manufacturing process and that they are able to predict the influence of manufacturing parameters (*e.g.* slurry formulation, drying rate, calendering pressure) on the LIB electrode microstructures. Furthermore, we have demonstrated 3D-resolved models receiving as an input such predicted microstructures and predicting their electrochemical performance. While the manufacturing models have been experimentally validated by us and that the performance model provided performance predictions upon discharge with order of magnitudes close to the experimental ones, a 1-to-1 quantitative comparison between the performance model predictions and experimental discharge curves was not yet explored by us. In this work we present an experimentally validated 3D-resolved electrochemical model of a NMC111-based electrode which reveals how slurry formulation and calendering degree affect the electrode performance. A series of electrodes with different formulations and calendering degrees were fabricated at the experimental level. Corresponding three-dimensional manufacturing models were built based on the same experimental manufacturing parameters to generate the digital counterparts of the experimental electrodes that were then used in the electrochemical model. The results of simulations and experiments were compared individually. Among the manufacturing parameters analyzed, we found that the major factors linking manufacturing parameters and electrode performance are the carbon and binder domain (CBD) distribution within the electrode volume, and the electrostatic potential difference between the electrode and the current collector. A well-connected electronic conductive network throughout the electrode is vital for ensuring full utilization of active material, and it was found that increasing calendering degree is effective in reducing interfacial impedance. This work uncovers, based on a dual modeling/experimental approach, the essence of how electrode manufacturing process takes effect on electrode performance by influencing its microstructure.

# 1. Introduction

Simulation research in the field of lithium ion batteries (LIBs) has progressed significantly in the last years. From the use of the so called equivalent electric circuit models to the homogeneous Newman model, until the most modern heterogeneous models taking into account explicit 3D representations of the electrode microstructure, the researchers are striving for revealing the intrinsic physics within LIBs[1]. The final goal of this task is providing guidelines for improving LIBs electrochemical performance, durability and safety. In this sense, fruitful results have been achieved. For instance, Lu *et al.* [2] made electrochemical 3D-resolved computational analysis based on X-ray tomography data of a NMC-based cathode. Comparisons were made regarding different electrode designs and calendering degrees, showing the complex interplay of various physics processes within the electrodes. De Lauri *et al.* [3] investigated in heat generation and lithium plating during fast charging based on stochastically generated electrode structures. The study suggested that optimizing electrode microstructure can be beneficial to reduce electrolyte transport resistance. Parmananda *et al.* [4] suggested that the homogeneity of particle morphology is important for graphite electrodes performance and safety by characterizing the electrode microstructure features through a multiscale approach coupling microstructure and mesoscale Finite Volume Method (FVM) models. The importance of various electrode features has also been discussed by researchers in the field, such as the effect of thickness, inner stress, deformation, particle shape, carbon-binder distribution, etc. [5–8] Most of the microstructure-based studies are done with digital structures that are either extracted from tomography data, or generated stochastically. Despite the ability of providing in-depth understanding, many simulation studies still lack of quantitative and practical guidance for the experimental preparation of electrodes because the manufacturing parameters were not taken as inputs of the models. LIB electrode manufacturing involves a series of processes, which can be roughly categorized in four main steps:

mixing, coating, drying and calendering. Within this workflow, a complex ensemble of parameters needs to be controlled at each step to meet the production requirements, such as material choice, slurry formulation, mixing time and rate, coating speed and comma gap, drying temperature and time, calendering pressure, etc. The choice of the parameters will have direct influence on the final performance of the electrodes. A possible pathway to account explicitly for manufacturing parameters is using a full digital replica of the entire fabrication process, which can allow to study digitally the correlation between manufacturing parameters and electrochemical performance.

The most critical aspect for this type of digital replica is the generation of the electrode microstructure. Many methodologies have been implemented to build digital electrodes being representative of real ones, like stochastic models, experimental imaging techniques, machine learning and physics-based models. Among all the approaches, physics-based model has been proven to be a good compromise between cost and accuracy for building digital twins of the electrode manufacturing process [9–13]. Generally speaking, this kind of model typically utilizes techniques such as discrete element method (DEM) to simulate the electrode particle ensemble and their evolution as a function of the manufacturing parameters. The interaction between particles are described by a set of forces or force fields, *i.e.*, mathematical equations linked to given particle-particle interactions, as for instance adhesive or mechanical forces. Our group's ARTISTIC project pioneered the digitalization with physics-based models of the entire key manufacturing process steps (slurry, drying, calendering, electrolyte infiltration and electrochemical performance)[14]. In our previous studies, we have made thorough discussions regarding the electrode manufacturing and electrochemical analysis models [15–17]. While our manufacturing models have been experimentally validated by us and that our performance models provided performance predictions upon discharge with order of magnitudes close to the experimental ones, a 1-to-1 quantitative comparison between the performance model predictions and experimental discharge curves for different manufacturing conditions was not yet explored by us. This comparison between the performance model predictions and the experimental discharge curves for different manufacturing parameters represents the main novelty of the present work. In particular, we analyzed the effect of electrode formulation and calendering degree on the discharge performance. In the following, we start by describing our research methodology, then we present and discuss our results, and finally we conclude and indicate future directions of our work.

## 2. Experiment

$LiNi_{0.33}Co_{0.33}Mn_{0.33}O_2$ (NMC111) is chosen as our target material. Three electrode formulations, accounting for different weight ratios of active material (AM), conductive carbon black and polyvinylidene fluoride (PVdF) binder, were studied: 85:9:6, 90:6:4 and 95:3:2. Here, the carbon to binder ratio was kept constant to 1.5 for each formulation. First, AM, carbon black and binder were premixed overnight. The materials were then stirred with N-methylpyrrolidone (NMP) using a Dispermat CV3-PLUS high-shear mixer for 2h at 25 °C. The solid contents of the slurries are: 46% for the 85wt% electrode, 55% for the 90wt% electrode and 67% for the 95wt% electrode. The slurry is then coated on aluminum foil (thickness of *ca.* 22 μm) with a comma-coater prototype-grade machine. Different coating comma gaps were chosen to achieve the desired mass loadings: 85 wt% electrode is 300 μm, 90 wt% electrode is 205 μm and 95 wt% electrode is 130 μm. NMC111 (average particle diameter = 5 ± 3 μm) was supplied by Umicore. C-NERGY™ super C65 carbon black (CB) was supplied by IMERYS. Solef™ 5130/1001. PVdF was purchased from Solvay and NMP from BASF.

After drying, the electrodes were calendered until compressing their initial thickness of 10% and 20%. Including the uncalendered electrodes, we analyzed three different electrodes for each formulation. Greater calendering degree were not considered because they would have increased the likelihood of particles cracking, which are not explicitly accounted in our calendering model used in this work. For convenience in the following, the samples are represented with their weight ratios of AM with suffix of calendering degree: for example, 85Cal0 stands for the electrode with formulation of 85:9:6 with calendering degree of 0% (uncalendered). The AM loading of all the electrodes was kept constant to *ca.* 15 mg cm$^{-2}$. The electrodes were then assembled with lithium metal reference electrodes and Celgard separator into 2032-type coin cells. The electrolyte used is LP30 which contains 1 mol/L LiPF$_6$. The solvent is 1:1 mixture of ethylene carbonate (EC) and dimethyl carbonate (DMC). Three coin cells for each condition were tested electrochemically to ensure the representativeness of the results. Galvanostatic charge-discharge cycling tests were then carried out for the cells at room temperature (RT) under 0.5C, 1C and 2C discharge rates (3 cycles per C-rate). The charge was performed, for each C-rate at 0.1C plus a constant voltage step (until current < 0.05 C) to ensure that each discharge started from the same state of charge. Before starting such an electrochemical protocol, a formation cycle (0.02 C in both charge and discharge) was performed. The potential window utilized ranged between 3.2 V and 4.2 V. Before cycling each cell, at least 6 hours were waited, in order to give the electrolyte enough time to fill as much electrode and separator pores as possible. In the following simulation work, 0.5C, 1C, and 2C experiment data were chosen as representative study cases.

## 3. Model description

The modeling workflow consists of two main parts. The first part is devoted to the generation of the electrode microstructures. The second part is devoted to the simulation of the electrochemical behavior of the electrodes upon discharge.

**3.1 Electrode microstructure generation**

The generation of the electrode microstructures was performed using coarse-grained molecular dynamics (CGMD)/Discrete Element Method (DEM) simulations using LAMMPS computational software. These simulations account for both the AM and CBD phases, but it does not distinguish explicitly between carbon and binder. Three formulations, corresponding to their experimental counterpart, were simulated, *i.e.*, 85:15 (358 AM and 21094 CBD particles), 90:10 (358 AM and 13281 CBD particles) and 95:5 (358 AM and 6648 CBD particles) weight percentages. The solid contents used are the same ones used for their experimental counterpart. To simulate the slurry and the drying process, we used our previously published CGMD models, in which the slurry was obtained after equilibration of the initially randomly generated structure in NPT environment at 298 K and 1 atm. The initial structure was generated to match the experimental composition and AM particle size distribution by placing AM and CBD particles in a simulation box big enough (several hundreds of µm per each Cartesian direction) to avoid significant overlap between particles at the beginning of the simulation. The exact ratio between x, y, and z sizes of the initial structure was adjusted to get, for each final electrode microstructure, an AM loading of *ca.* 15 mg cm$^{-2}$ like in the experimental counterparts. The drying was performed in the NVT environment at 363 K by decreasing the CBD size (containing not only the carbon and binder, but also the solvent at the slurry phase) in order to mimic the solvent removal. For both slurry and drying simulations periodic boundary conditions (PBCs) were applied in all the directions, to enhance the representativeness of the simulated microstructures. The readers can refer to our previous work for more details about the simulation of the slurry and the drying process [15,17,18].

Concerning the calendering step, it was found that shifting from the LAMMPS framework to the LIGGHTS one, as in one of our previous publication [18], can lead to particle reorganization in certain scenarios (in terms of AM wt.% and electrode thickness). To avoid this, calendering DEM simulation here is also performed in LAMMPS. To simulate the calendering process, two planes were added at the top and the bottom of the dried electrode microstructures, representing the calendering roll and the current collector, respectively. Both planes interact with the particles through the Granular Hertz force fields. During the calendering, the top plane is moved down until reaching the desired compression degree (here a compression of either 10 % or 20 % of the initial electrode thickness). In the calendering simulation, PBCs were applied on x and y directions only, while the z direction was considered as non-periodic. The force field parameter values and codes used are the same adopted in our ARTISTIC online calculator [20].

### 3.2 Electrochemical model

The electrode microstructures generated from the CGMD/DEM modeling workflow were then voxelized with resolution of 0.5 µm, by using INNOV [21,22] a home-developed software which is able to build meshes for finite element method (FEM) calculations. Commercial software COMSOL Multiphysics® was used as our FEM calculation platform.

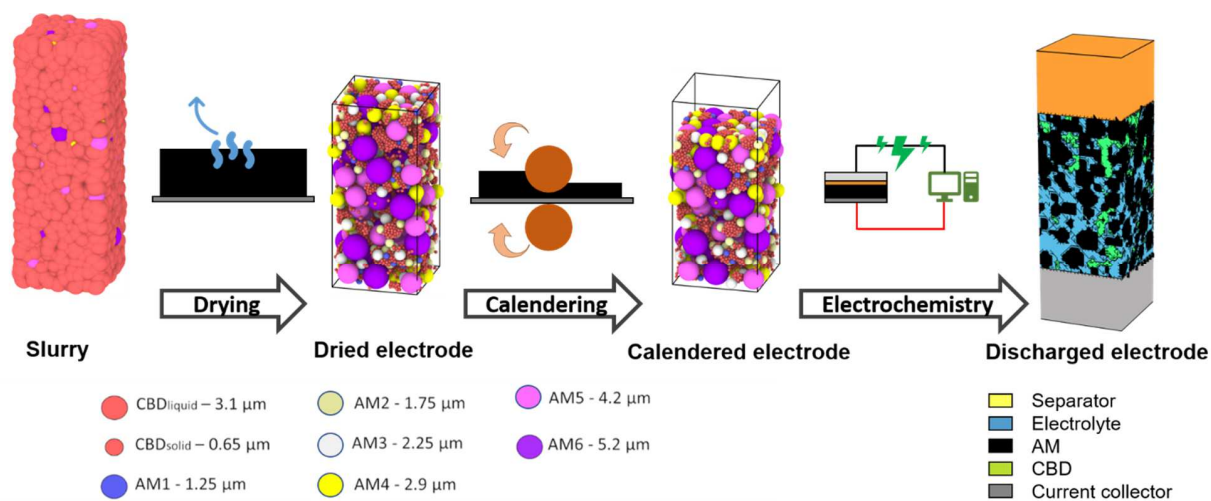

**Figure 1**: Illustration of the modeling workflow of 95Cal20 as an example. From slurry generation to calendered electrode, the model is built and run in LAMMPS. The output is then meshed with INNOV and imported in to COMSOL. The side length of the current collector is 35 μm. The mesh consists of 849843 elements. The value can vary for different electrodes because of varying sizes.

The computational domain of our electrochemical model consists of five phases: AM, CBD, electrolyte, separator and current collector (CC). Figure 1 illustrates the geometry of 85Cal0 as an example. All the equations and parameters utilized are listed in Table S1. Equilibrium potential of NMC material is taken from the average profile of three parallel experimental results of 0.02C discharge of the 90Cal20 electrode. The window of degree of lithiation (DOL) of NMC is also calculated based on the ratio of experimental reversible capacity and theoretical maximum Li content. Assuming NMC is fully lithiated at the end of discharge, the DOL at fully charged state (electrode potential at 4.2 V) is 0.45925. Solid diffusion coefficient in NMC domain as a function of DOL comes from the work of Wu *et al.* [23]. The curve is further modified by adding a correction term: $(2^{-15} \cdot DOL - 1^{-15})$ m$^2$ s$^{-1}$ in order to increase the diffusion coefficient at the vicinity of full lithiation state for better fitting. According to previous studies [24–27], the lithium diffusivity in NMC111 lies in the magnitude of ~10$^{-15}$ m$^2$/s, therefore we believe that this correction is reasonable.

CBD is considered as homogeneous porous medium in the model with a porosity of 27% [28]. Based on the study of Trembacki *et al.* [29], due to the high tortuosity feature within CBD phase, the transport process is much lower than in the electrolyte phase. Therefore, the ionic diffusivity and conductivity are modified to 5% of the bulk value. The expression of reaction exchange current density included the CBD porosity at the interface with the NMC111 particles to take into consideration the limited exposure of the NMC particles surface to the electrolyte.

**4. Result and discussion**

Our 4D-resolved electrochemical model is validated by comparing the simulation results with experimental discharge profiles, as illustrated in Figure 2. The key findings from the analysis of the experimental dataset are as it follows:

- Generally speaking, the formulation with 90 wt% of AM exhibits the best electrochemical performance, in which 90Cal20 has the highest specific capacity;
- For each formulation, the specific capacity of the electrode increases with higher calendering degree;
- The capacity of the formulation with 85wt% of AM has limited degradation comparing to 90 wt% cases. Apart from that, the potential plateaus are also evidently lower;
- Formulation with 95 wt% of AM shows severe deterioration of electrochemical performance. For example, 95Cal20 shows a specific capacity of 88.09 mAh $g^{-1}$ at the end of 2C discharge, while for 90Cal20 case, the value is 122.27 mAh $g^{-1}$.

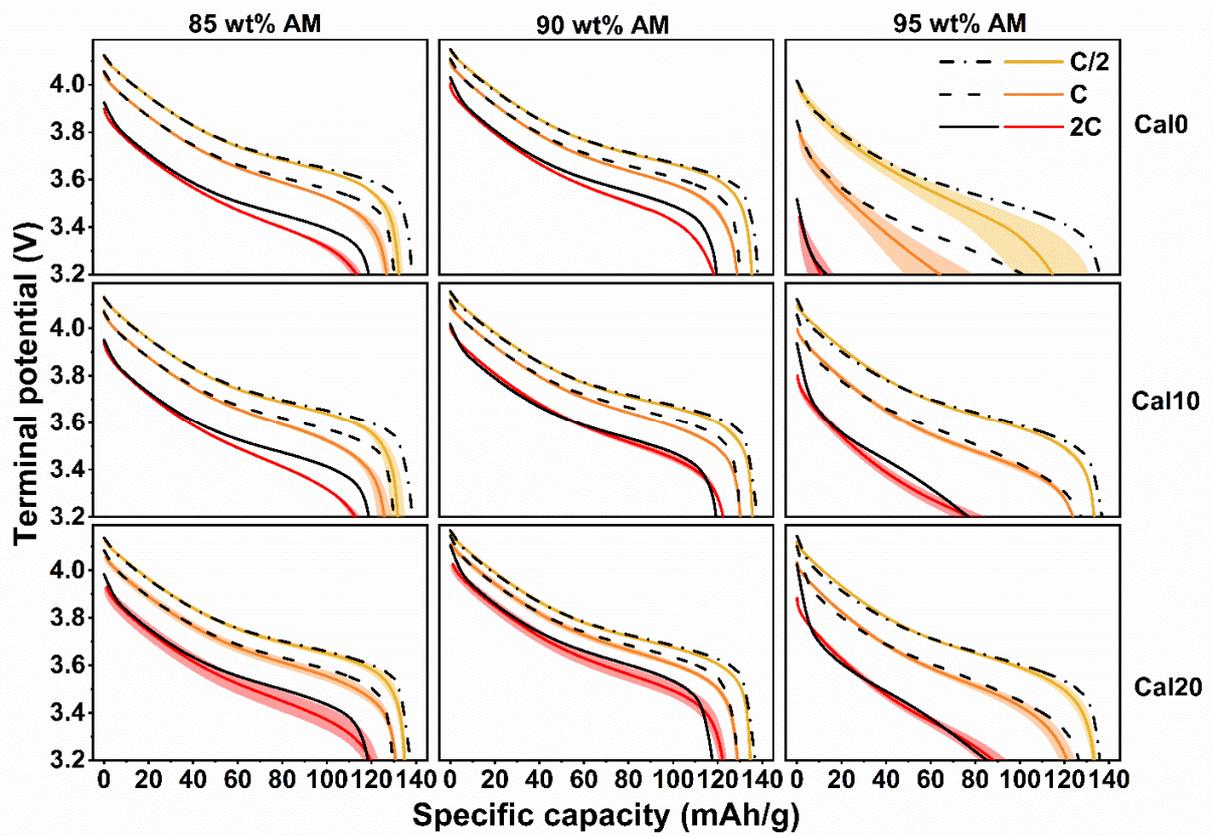

**Figure 2**: Comparison of experimental discharge profiles at 0.5C, 1C and 2C with the simulation results. Experimental data are presented as yellow, orange and red lines where solid line is the average profile of the 3 parallel tests, and the area represents the standard deviation. Simulation results are presented as black solid, dashed, and dotted lines for 0.5C, 1C, and 2C, respectively. Columns and rows correspond to formulations and calendering degrees, respectively.

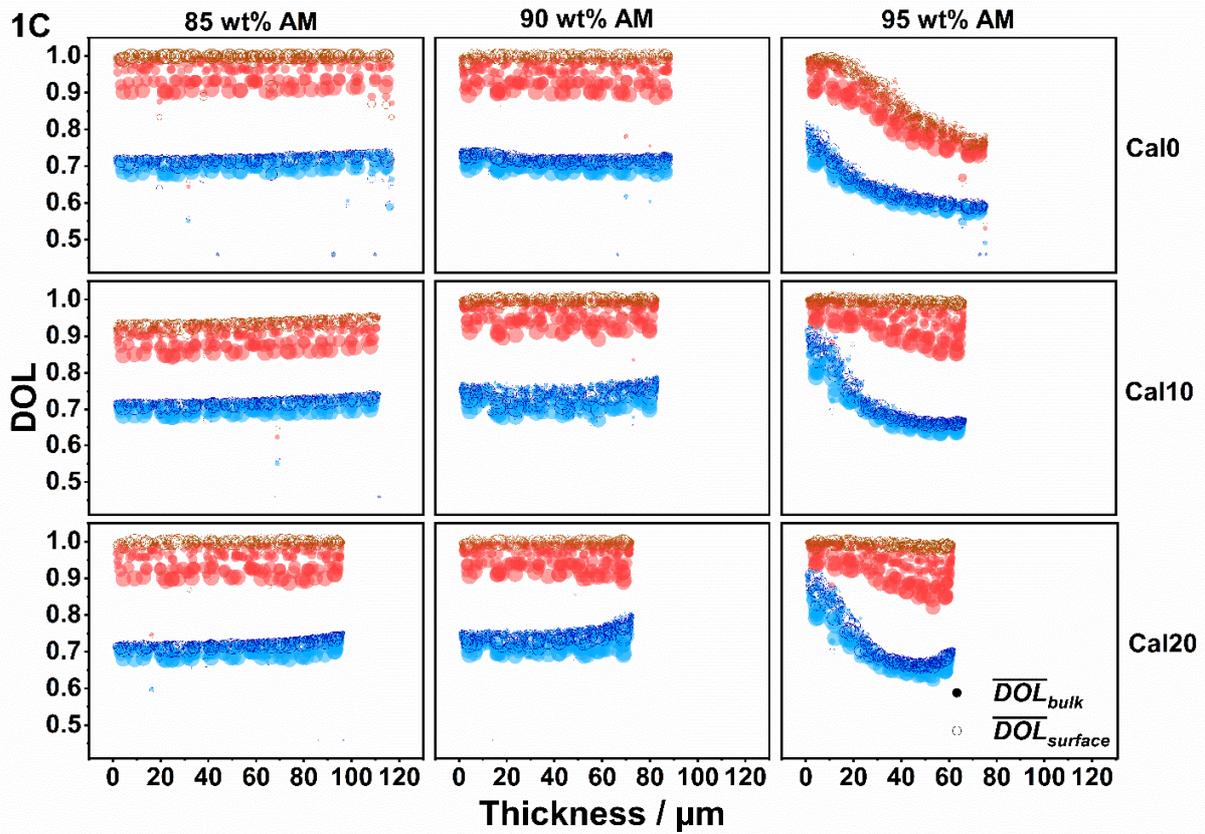

**Figure 3**: Illustration of DOL state of each particle in the model of 1C discharge. X axis is the position of the particles' centroids in the thickness direction. The thickness of 0 µm refers to the interface of the electrode with the CC. DOL of 50% discharge state is marked as blue and 100% discharge state is red. Solid points represent $\overline{DOL}_{bulk}$ and hollow points represent $\overline{DOL}_{surface}$.

The discussions in the following will be centered around the above points as a function of our analysis of the simulation results. For the convenience of discussion, 90Cal20 electrode, which exhibits the best performance, is taken as the reference case and all the other electrodes are analyzed by comparing to it. To visualize the electrochemical state of the system, the DOL state of each particle in the electrode and overpotential profiles are summarized in Figure 3 and Figure S2-S3. DOL of a particle is categorized into two parts: average value of bulk and surface, in a way that the effect of solid phase diffusion can be visualized. Overpotentials are illustrated as Li⁺ concentration difference and solid potential difference on the separator side and the CC side, and the average faradaic current density on the active material surface. Briefly, three factors were found to be major leads to the discharge profile differences, including Li diffusion resistance within the NMC particles, electronic conductive network and contact resistance between the electrode and the CC, which are singularly discussed in more details below

**4.1 Diffusion within the NMC phase**

Within the different NMC materials, NMC111 has relatively low diffusivity in comparison to high nickel content materials. Consistently, our model shows that the diffusion resistance of Li in the NMC phase is the major limiting factor for the different C-rates. To intuitively characterize the influence of the overpotentials, the average DOL value in each particle ($\overline{DOL}_{bulk}$) and on each particle surface ($\overline{DOL}_{surface}$) is plotted in Figure 3 and Figure S2-S3. DOL of 50% discharge state is marked as blue and 100% discharge state is red. The sizes of the plot points are proportional to the size of particles. Generally speaking, for 85wt% and 90wt% cases, $\overline{DOL}_{surface}$ is always higher than $\overline{DOL}_{bulk}$ due to diffusion resistance within solid phase. At 100% discharge state, $\overline{DOL}_{surface}$ reaches 1 while $\overline{DOL}_{bulk}$ is lower and dispersed. Small particles have higher DOL than large particles due to the shorter diffusion path within the particles. As C-rate increases, $\overline{DOL}_{bulk}$ is more dispersed, causing larger capacity degradation. At 50% discharge state, the electrode shows a significant higher DOL at the separator side compared to the current collector side. This is due to the Li$^+$ concentration distribution throughout the electrode thickness. Also since the DOL of particles are at a relatively low state, which means that the diffusivity in the solid phase is relatively high, therefore the ionic transport in electrolyte takes considerable effect on the DOL distribution. However, as the DOL increases, the solid phase diffusivity decreases to the extent where it becomes the dominating factor. Thus, the DOL profile does not show distribution in the thickness direction at 100% discharge state. Therefore, the liquid phase mass transport resistance is not vital in the scope of our study cases. The DOL distribution of 95 wt% electrodes exhibit a completely different pattern compared to 85 wt% and 90 wt%. The solid phase diffusion still makes the DOL value disperse but not to the extent of such capacity deterioration as shown in Figure 2. The limiting factor of the 95 wt% electrodes is changed to the electronic conductivity which will be discussed in the following section.

**4.2 Conductive network**

Carbon black and binder are important components of an electrode. This composite additive does not only maintain the mechanical integrity of the electrode, but also ensures good electronic conduction by forming a conductive network. The effective electronic conductivity of the system is attributed to two aspects: intrinsic conductivity of the CBD and morphology of the CBD network. The intrinsic conductivity is affected by the ratio of carbon black and binder, which is not in the scope of this study, and by the compression degree. There have been researches showing that the CBD conductivity increases while compressing. Trembacki *et al.* [29] reported a correlation between CBD conductivity and volumetric strain, which is fitted to a linear expression and adopted in our study with a factor of 10 (unit: S/cm):

$$\sigma_{cbd} = -173.967\epsilon_{v,m} + 0.1593 \tag{1}$$

In this study, we do not resolve the heterogeneity of the CBD strain. Instead, we considered that the CBD deformation is homogeneous for simplicity. $\epsilon_{v,m}$ is calculated by the following equation:

$$\epsilon_{v,m} = \frac{V_{cbd,m}}{V_{cbd,uncal}} - 1 \tag{2}$$

$V_{cbd,m}$ is the volume of CBD phase. $V_{cbd,uncal}$ is the volume of CBD phase of the uncalendered case. The volume data is extracted from the voxelized CGMD/DEM microstructure. In the model framework, the volumetric change of CBD phase is originated from the force field description of the discrete system. In particular, after calendering, due to the inner stress, the CBD particles will have a certain degree of overlap, which is considered as corresponding strain. Table S3 summarized the volumetric strains. It shows the trend that the absolute value of deformation increases as compression rate increases.

The CBD morphology is mostly determined by its formulation, and characterized by the degree of continuity of the CBD phase. Since the conductivity of the NMC111 is about three order of magnitudes lower than the one of CBD, the electron transfer within the electrode is highly dependent on the integrity of the CBD phase. The results of 95wt% electrodes are shown as good examples of the effect of CBD morphology. Figure 3 and Figure S2-S3 shows that 95wt% electrodes always show a higher DOL at the CC side, especially for high C-rate. This phenomenon is attributed to the bad connection of the CBD phase due to its low content, which causes the particles on separator to lack fast access to electrons. Figure S5 shows the solid phase potential drop from separator side to CC side. For cases with 85wt% and 90wt% of AM, the content of CBD is high enough to form a complete connection, hence potential drop in solid phase is relatively low. For cases with 95wt% of AM, the potential drop is one magnitude higher. Figure 4(a) intuitively illustrates the potential profile of the CBD phase of 85Cal20, 90Cal20 and 95Cal20 at the end of 1C discharge. We noticed that CBD phase is dispersed into agglomerates instead of forming a continuous domain throughout electrode like 85 wt% and 90 wt% cases, therefore 95Cal20 has a significantly large potential drop through thickness direction, while 85Cal20 and 90Cal20 hardly any potential distribution that could be recognized under the same legend range.

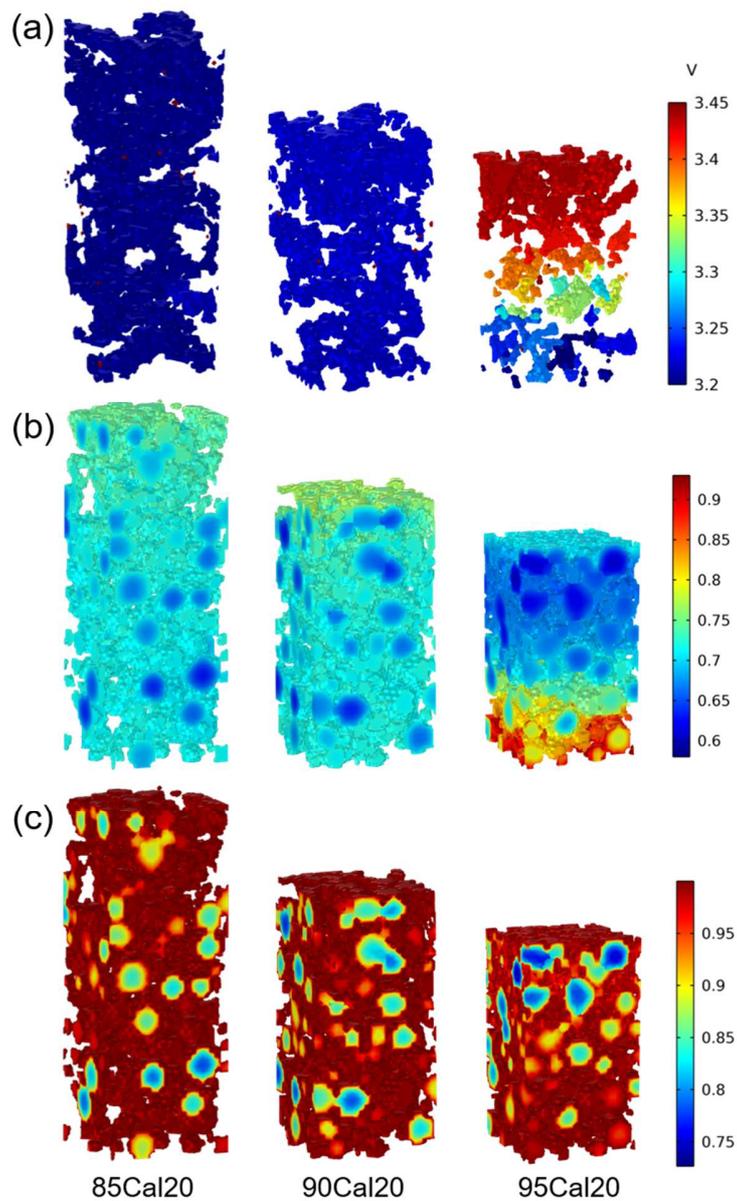

**Figure 4**: Illustration of 85Cal20, 90Cal20 and 95Cal20 electrodes features of (a) solid potential distribution within CBD phase at the end of 1C discharge, (b) DOL within AM phase at midway (1400s) of 1C discharge and (c) at the end of 1C discharge.

To quantitatively characterize the effect of CBD morphology, the electrode effective conductivities are calculated and summarized in Figure 5(a). 95 wt% electrodes have generally lower conductivities. As CBD content increases, the electrode conductivity also increases. The calendering process is also beneficial to improving electron conduction, because compressed CBD phase has higher conductivity.

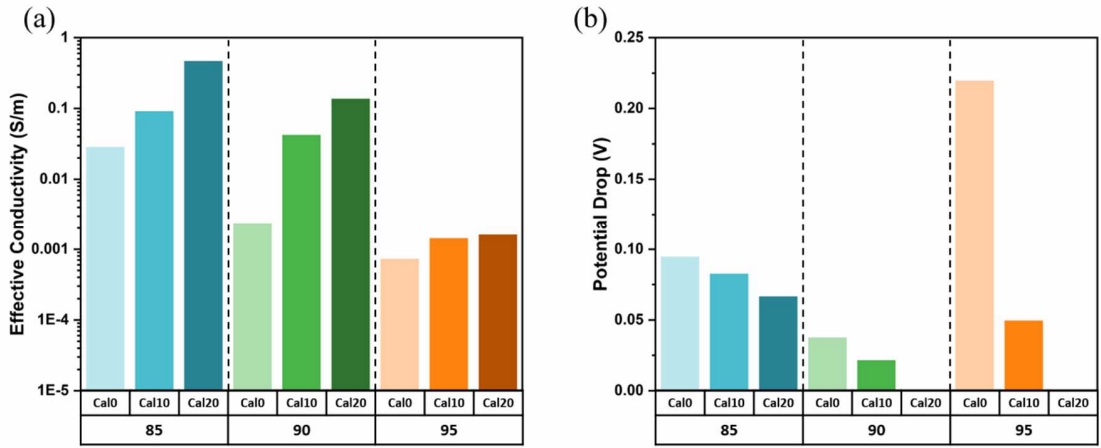

**Figure 5**: (a) Calculated electrode effective conductivity based on the manufacturing simulation-generated geometry. (b) Potential drop at the interface of the electrode with the CC at 1C discharge.

**4.3 Interfacial resistance**

The LIB simulation community has been concentrating the efforts on explaining microstructure morphology, which derives sub-questions about optimization of porosity, tortuosity factor, particle size, thickness and so on. However, the contact resistance between the electrode and the CC is a feature that truly exists but that is seldomly studied. Experimental studies have shown that contact resistance is one of the major sources of battery impedance. Gaberscek *et al*. [30] pointed out that the contact resistance between the electrode and the CC is the main source of high frequency impedance in electrochemical impedance spectroscopy (EIS) experiments. Type of CC metal and applied pressure make a significant difference to the high frequency impedance value. Nara *et al*. [31] quantified the interfacial resistance between the CC and the electrode to be around 0.9 Ω through EIS, and further demonstrated that carbon coated CC can significantly reduce the interfacial resistance. Pritzl *et al.* [32] found that the contact resistance between the electrode and the CC is noticeably reduced with calendering. Their study also shows that the contact resistance increases during cycling, which enriches the community's perception of the mechanisms behind the LIB cell degradation. Based on these studies, we believe that it is vital to include in the electrochemical models this interfacial resistance for better interpretation of the system. Therefore, in our model a thin layer of electrode is defined as the interface of CC with electrode using the following equation:

$$\boldsymbol{n} \cdot \boldsymbol{i}_s = \frac{\Delta \phi_s}{R_{interface}} \quad (3)$$

In the equation above $\boldsymbol{n}$ is the normal vector at the boundary, $\boldsymbol{i}_s$ is the electronic current density, $R_{interface}$ is the contact resistance with unit of Ω·cm$^2$, and $\Delta \phi_s$ is the potential drop at the boundary. $R_{interface}$ is acquired by fitting to the experimental results.

The results show that calendering process has significant effect on reducing electrode-CC contact resistance apart from increasing electrode effective conductivity in the scope of our performance model. Figure 5(b) illustrates the potential drop in the interface between the CC and the electrode. For each formulation, the potential drop decreases with calendering due to a better interfacial contact. We can also notice that the interface potential drop of 95 wt% is dramatically higher than other formulations. This indicates that the contact resistance is not only related to the calendering degree, but also to the formulation. It also shows for 95 wt% the major effect of calendering process is reducing electrode-CC contact resistance. We can see from Figure 5(a) that regarding 95 wt% electrodes, the effective conductivity remains at the level of 2~3 magnitudes lower than electrodes of other compositions. Figure S5 shows that the values of potential drop along the thickness direction of 95 wt% electrodes are always 1 magnitude higher comparing to other compositions. Even though a proper quantitative estimation of its impact is difficult, these simulation results emphasize the importance of considering interface contact resistance in electrochemical simulations.

**5. Conclusion**

In this work, we have presented a physics-based 3D electrochemical model of a NMC111 electrode, combined with 3D manufacturing simulations, which was calibrated with a series of experimental results of various electrode formulations and calendering degrees. Our study investigated the main phenomenon causing different electrochemical performances as a function of C-rate, electrode formulation and calendering degree. The capacity degradation at high C-rates is attributed to the low diffusivity of Li in the NMC111 AM. Larger particles tend to have lower DOL at the end of discharge causing insufficient utilization of AM. Electrode formulation significantly influences the conductive network integrity. Low content of carbon black and binder causes CBD phase to become discontinuous, resulting in poor electrode electronic conductivity. Therefore, choosing the right formulation is vital to have sufficient CBD, and sufficiently interconnected, to retain high electronic conductivity. The optimum point in our study case is between 90 wt% and 95 wt% of AM in the formulation. Increasing the calendering degree is beneficial for obtaining better electrochemical performance, because CBD has higher conductivity due to the more compact state. Also, more importantly, increasing the calendering degree decreases the electrostatic potential difference between the electrode and CC, which implies a better contact between CBD and CC. Our results point out that the consideration of the interfacial resistance between CBD and CC in relation to the manufacturing process is vital to be included and investigated in LIB cell performance simulations. Future directions of this work include the integration of our electrochemical model in the ARTISTIC project online calculator[33] as a free online service for battery stakeholders from academia and industry.

**Acknowledgements**

C.L, T.L., A.C.N. J.X and A.A.F. acknowledge the European Union's Horizon 2020 research and innovation program for the funding support through the European Research Council (grant agreement 772873, "ARTISTIC" project). A.A.F. acknowledges Institut Universitaire de France for the support. The authors acknowledge the MatriCS HPC platform from Université de Picardie-Jules Verne for the support and for hosting and managing the ARTISTIC dedicated nodes used for the calculations reported in this manuscript.